\documentclass[twocolumn,floatfix,showpacs,preprintnumbers,amsmath,amssymb]{revtex4}

\usepackage{dcolumn}
\usepackage{bm}
\usepackage{graphicx,psfrag}
\usepackage{epsfig}

\newcommand{\be}{\begin{equation}}
\newcommand{\ee}{\end{equation}}
\newcommand{\bear}{\begin{eqnarray}}
\newcommand{\eear}{\end{eqnarray}} \newcommand{\ba}{\begin{array}}
\newcommand{\ea}{\end{array}}

\def\beq{\begin{equation}}
\def\eeq#1{\label{#1}\end{equation}}
\def\eeqn{\end{equation}}
\def\eeq{\end{equation}}
\def\beqa{\begin{eqnarray}}
\def\eeqa#1{\label{#1}\end{eqnarray}}
\def\eeqan{\end{eqnarray}}

\newcommand\iden{\leavevmode\hbox{\small1\normalsize\kern-.33em1}}

\def\W3{W_H^3}

\begin{document}

\preprint{FERMILAB-PUB-09-250-A}

\title{Supersymmetric and Kaluza-Klein Particles Multiple Scattering in the Earth}
\author{Ivone~F.~M.~Albuquerque}
\affiliation{Center for Particle Astrophysics, Fermi National Accelerator Laboratory, Batavia, IL, 60510 
and Instituto de F\'isica, Universidade de S\~ao Paulo, S\~ao Paulo, Brazil}
\author{Spencer~R.~Klein}
\affiliation{Lawrence Berkeley Laboratory, CA, 94720 and University of California, Berkeley, CA, 94720}
\vspace*{0.3cm}


\begin{abstract}
Neutrino telescopes with cubic kilometer
volume have the potential 
to discover new particles. 
Among them are next to lightest supersymmetric (NLSPs) and next to lightest 
Kaluza-Klein (NLKPs) particles. 
Two NLSPs or NLKPs will transverse the detector simultaneously
producing parallel charged tracks. The track separation inside the detector 
can be a few hundred meters. 
As these particles  might propagate a few thousand kilometers before reaching the detector, multiple 
scattering could enhance the 
pair separation at the detector. We find that the multiple scattering will alter  
the separation distribution enough to increase
the number of NLKP pairs separated by more than 100~meters 
(a reasonable experimental cut) by up to 46\% depending on the NLKP mass.
Vertical upcoming NLSPs will have their separation increased by 24\% due to 
multiple scattering.
\end{abstract}

\maketitle

\section{Introduction}
\label{sec:intro} 
One of the contemporary questions in particle physics regards extensions of the
standard model at higher energy scales. 
New physics should appear around TeV energy scales. 
Large telescopes for high-energy astrophysical neutrinos can complement colliders 
in searching for signatures of extension models.  In some models, their sensitivity
can extend beyond the reach of the LHC.
More specifically it was shown that 
neutrino telescopes can directly detect next
to lightest supersymmetric particles (NLSP) \cite{abc,abcd} and next to lightest 
Kaluza-Klein (KK) particles (NLKP) \cite{abkn}. In scenarios were the lightest
supersymmetric (KK) particle is the gravitino (KK mode of the graviton), 
charged right-handed sleptons (KK lepton modes) will be the NLSP (NLKP).
In the supersymmetric scenario, neutrino telescopes 
searches will complement collider searches
in the determination of the supersymmetry breaking scale \cite{abc}.

Searches for both NLSPs and NLKPs complement dark matter searches.
The lightest particle in each of these scenarios are dark matter candidates.
Both the gravitino and the KK mode of the graviton cannot be detected, at least
with the current technologies. Therefore NLSP (NLKP) searches also 
probe dark matter, albeit indirectly. Other supersymmetry scenarios 
with charged leptons as NLSPs to be 
probed by neutrino telescopes were considered 
by \cite{madrid}. These assume 
scenarios where either the neutralino or a super weakly massive interacting 
particle is the LSP. 

High energy (above ~$10^5$~GeV) neutrino interactions 
in the Earth may produce supersymmetric or KK particles. These will go 
through a decay chain \cite{abc,abkn} which will always lead to
the production of two NLSPs or NLKPs. These next to lightest particles will
propagate through the Earth, eventually reaching a neutrino telescope. 
 Due to their heavy mass, their specific energy loss (dE/dx)
as they propagate through the earth is much smaller \cite{abcd,ina}
than the loss by standard leptons. 
For this reason NLSPs and NLKPs may have a range of thousands of kilometers, while even
the most energetic muons have a range of a few tens of kilometers.
The smaller energy loss compensates the small
NLSP and NLKP production cross sections.

One signature  for these new particles is two charged
tracks transversing the detector \cite{abcd,abkn}. Here we investigate the NLSP and NLKP multiple 
scattering while propagating through the Earth.  
We consider the simulated 
distribution of their production point and energy loss as described in 
\cite{abcd,abkn}.
The separation distance between the two charged tracks in the detector will be a
powerful discriminating observable. 
It is an excellent
parameter to separate NSLPs or NLKPs from the main background, which 
consists of dimuon events. As a preview to our conclusion, we 
note that multiple scattering
will increase the number of NKLP pairs separated by more than 100~meters 
up to 46\% depending on its mass. Although the NLSP pair separation increase is
smaller (3\%) the multiple scatter increases the number of vertical upcoming NLSPs
by 24\%.

In the next section we describe how we account for multiple scattering
and energy loss. Then we show our results, followed by a discussion on the 
detectability of these particle. Our conclusions follow in the last 
section. 

\section{Multiple Scattering}
\label{sec:ms} 

The formulae for multiple scattering are well known; the scattering is independent of the
particle mass.   The angular deviation by a singly charged relativistic particle with momentum $p$
passing through a slab with thickness $L$ made out of a material with radiation length $X_0$ in space
(i.e., the total angular deviation, not that in a plane)
is well described by a Gaussian distribution \cite{pdg}.
The scattering angle 
in a given plane $\theta_0$ is defined as
\begin{equation}
\theta_0 = \frac{\rm 13.6\ MeV/c}{p} \sqrt{\frac{L}{X_0}}.
\label{eq:tht0}
\end{equation}

The radiation length, $X_0$ is calculated using Dahl's fit given in Equation~27.22 of \cite{pdg} to the 
data:

\beq
X_0 \;=\; \frac{716.4\ {\rm g\ cm^{-2}} A}{Z (Z+1) \ln(287/\sqrt{Z})}
\label{eq:radl}
\eeq
where $A$ and $Z$ are the atomic mass and number of the target. This equation
agrees with Tsai's \cite{tsai} values for Earth targets to better than 2.5\%. 

Since the earth is inhomogeneous, we use the density profile described in 
\cite{quigg}. This profile divides the Earth in ten layers.
The Earth structure, mean atomic number $\langle Z\rangle$, mass $\langle A\rangle$
of each layer 
depend on if it belongs to the crust, the mantle or the core of 
the Earth. These together with the mean density $\langle\rho\rangle$ and the 
radiation length $X_0$ are given in Table~\ref{tab:lay}.

\begin{table}
\begin{ruledtabular}
\begin{tabular}{ccccccc}
Layer & Structure & Height & $<Z>$ & $<A>$ &$<\rho>$ & $X_0$ \\
\# & & (km) & & & (g/cm$^3$) & (cm)\\
\hline
1 & Core & 0  & 25.6 & 54.8 & 12.93 & 1.10 \\ 
2 & Core & 1221.5 & 25.6 & 54.8 & 11.03 & 1.29 \\ 
3 & Mantle & 3480 & 27 & 55 & 4.97 & 2.56 \\
4 & Mantle & 5701 & 27 & 55 & 3.98 & 3.21 \\
5 & Mantle & 5771 & 27 & 55 & 3.85 & 3.33 \\
6 & Mantle & 5971 & 27 & 55 & 3.49 & 3.67 \\
7 & Mantle & 6151 & 27 & 55 & 3.37 & 3.80 \\
8 & Crust & 6346.6 & 12 & 24 & 2.90 & 8.66 \\
9 & Crust & 6356 & 12 & 24 & 2.60 & 9.66 \\
10 & Crust & 6368 & 12 & 24 & 1.02 & 24.63
\end{tabular}
\end{ruledtabular}
\caption{\label{tab:lay} Earth structure, mean atomic number $<Z>$, 
mass $<A>$, density $<\rho>$ and radiation 
length for each of the Earth's layers. The layers are as in the 
density profile given in \cite{quigg}. The height corresponds to the distance
from the center of the Earth to the beginning of the layer. Layer 1 starts at 
the center of the Earth. The top of the $10^{\rm th}$ layer is at the surface
of the Earth (6378 Km). The values for the crust, mantle and
core are given respectively in \cite{crust,mantle,core}.}
\end{table}

We calculate NLSP/NLKP energy loss following \cite{abcd,abkn}, and assuming that 
it is smooth.  The stochastic nature of the energy loss might broaden the distributions 
slightly, but should not be a significant factor here. 

We calculate the NLSP/NLKP divergence from the original neutrino trajectory.  The deviation due to the opening angle of the particle pair has been previously considered.  Here, we consider the additional deviation due to the multiple scattering of the NLSP pairs.

For each layer, the mean scattering angle is calculated from 
Eq.~\ref{eq:tht0}, and, from that angle, the deviation from the neutrino 
trajectory is projected to a surface detector.  The average deviation from 
the neutrino trajectory by a NLSP/NKLP particle due to scattering in layer $i$ is
$d_i = \sqrt{2}D\theta_0$, where $D$ is the distance from the middle of the layer to the surface; the $\sqrt{2}$ accounts for scattering in the $\vec{x}$ and $\vec{y}$ directions perpendicular to the direction of travel.

As in Ref.~\cite{abcd}, we integrate over different production points in the earth.  

The deviation in each layer is treated independently.  Since the angles are independent, the deviations are added in quadrature. As we neglect the small
initial energy difference between the two particles, and as they
scatter independently, their deviations are also added in quadrature and
determine the total separation due to multiple scattering. 
This separation is then added in quadrature (because the angles are uncorrelated) with the separation due to the initial opening angle.

Three factors affect the contribution to the multiple scattering from each layer.  The inner layers are the densest, and also have the longest lever arm, increasing their contribution.  However, the NLSP also have the higher energies while they are near the production point, decreasing the scattering.  Overall, the inner layers still make the largest contribution to the scattering - the lever arm wins out.

We also considered the effect of magnetic fields on the separation;  since the two NLSP have opposite charges, they are bent in opposite directions by the earth's magnetic field.  However, even after accounting for the increased field strength deep in the earth \cite{bfield}, this does not contribute significantly to the scattering.

\section{Results}
\label{sec:res} 

The analysis of NLKP detection by neutrino observatories \cite{abkn}
considered three mass values of 300, 600 and 900~GeV. 
The two NLKP track separation in the detector is the only discriminating
parameter for the highest mass considered although not the only one
for NLKP masses below 600~GeV. The background consists of
dimuon events, mostly from charm decay.

The NLKP energy loss depends on
its mass. The larger the mass the less its energy will degrade while the 
particle propagates through the Earth. This means that heavier NLKPs
can be produced farther from the detector. On one hand this 
enhances the multiple scattering,
since the particle will have 
a longer lever arm. However on the other hand the NLKP 
enters each layer with larger energy, reducing the scattering. 

\begin{figure}[tb]
\includegraphics[scale=0.4]{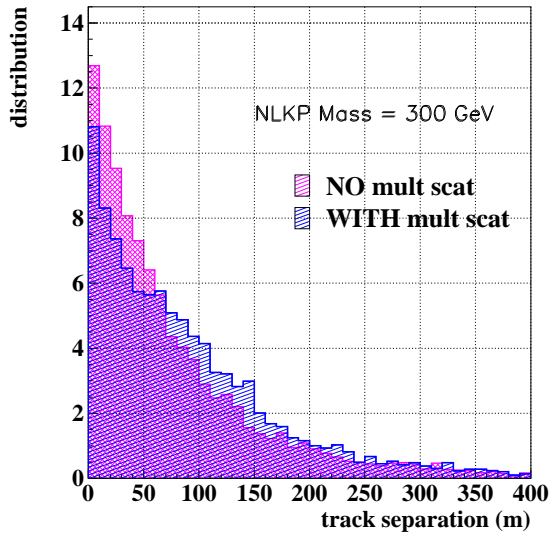}
\vspace*{-1.75cm}
\caption{\label{fig:NLKP3} Distribution of 300~GeV NLKP track separation 
distance in the detector. The separation is
shown with and without multiple scattering as labeled and both cases
have the same number of events.}
\end{figure}

\begin{figure}[tb]
\includegraphics[scale=0.4]{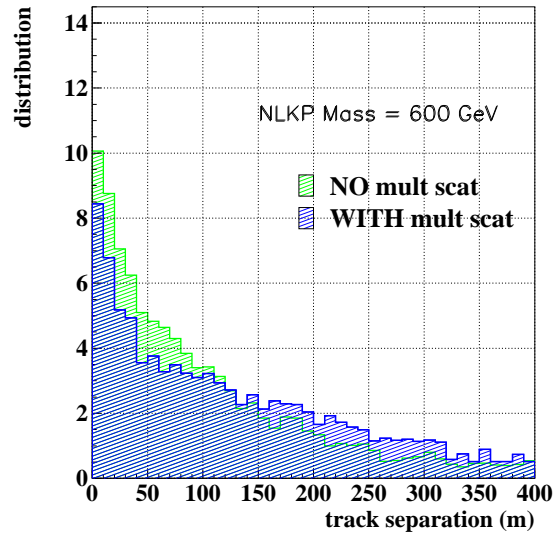}
\vspace*{-3.cm}

\includegraphics[scale=0.4]{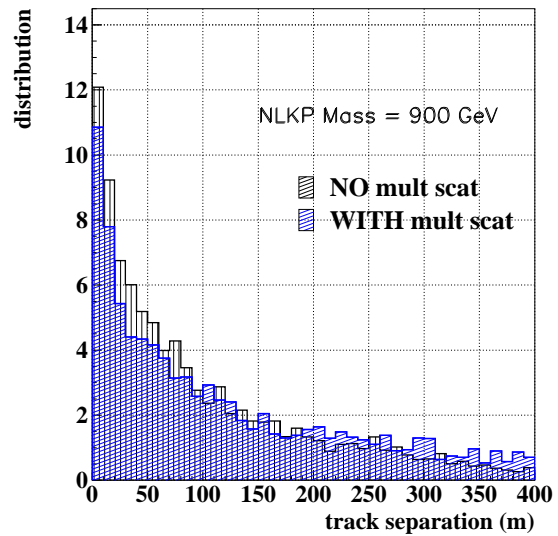}
\vspace*{-1.5cm}
\caption{\label{fig:NLKP9} Same as Figure~\ref{fig:NLKP3} but for 600
and 900~GeV NLKPs.}
\end{figure}

Figures~\ref{fig:NLKP3}~and~\ref{fig:NLKP9} show the effect of multiple scattering
on the NLKP pair separation in a detector positioned at the same depth in the Earth
as the IceCube detector (centered at 1950~m deep); the reduced scattering in the lower-$Z$ ice
is neglected. 
The 
separation is shown for three values of the
NLKP mass. The separation distributions are based on
30,000 simulated events. The separation due to the initial opening angle is 
described in \cite{abkn}. The multiple scatter
enhancement to the separation is described in the previous
section.

Multiple scattering has the largest effect on the separation distance for
NLKPs with a 
300~GeV mass. The number of events per km$^2$ per year, at a depth
equivalent to the IceCube detector with more than 100~m separation, rises
from 137 to 200 events when multiple scattering is included, a 46\% increase. 
For 600 and 900~GeV NLKPs, the corresponding increases are
30\% and of 21\% respectively. The background 
will be separated at most by 100 meters, as can be seen in Figure~7 of
\cite{abcd}. As described in the next section, 100~m
is 
the average detectable separation in km$^3$ telescopes.

NLSPs are produced from neutrino interactions in the Earth. This interaction
produces a charged left-handed slepton and
a squark as described in Ref.~\cite{abc,abcd}. These particles decay immediately
producing the two 150~GeV NLSPs. Here we show
results for
a 300~GeV squark. 
Multiple scattering has a similar effect on 600 and 900~GeV squarks.

\begin{figure}[tb]
\includegraphics[scale=0.4]{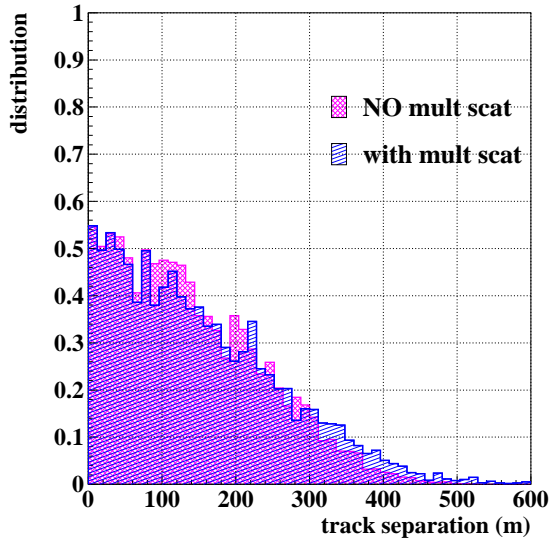}
\vspace*{-1.75cm}
\caption{\label{fig:NLSPsep} Track separation distance for NLSP pair events, 
in scenarios 
with a 300~GeV/c$^2$ squark mass, with and without multiple scattering.  
The two curves have the same number of events. The multiple scattering
enhancement on the separation distance is about the same for 600 and
900~GeV/c$^2$ squark masses.}
\end{figure}

Figure~\ref{fig:NLSPsep} shows the track separation between two NLSPs going
through the detector. The distributions show the separation due to the 
initial opening angle of the pair with and without multiple scattering.
The separation distance between the two NLSP 
is the main observable to discriminate it from the dimuon background.

The separation distributions shown in Figure~\ref{fig:NLSPsep} are based on
30,000 simulated events as described in Ref. \cite{abcd}. The darker (blue) histogram
represents the separation without multiple scattering while the lighter
(pink) the separation with multiple scattering. 

The multiple scatter effect in this case is weaker than
for NLKPs.
As the NLSP is lighter than NLKPs, they are produced closer to the detector
than NLKPs. 
The number of NLSP pairs with more than 100 meter separation will increase 
by 3\%. However multiple scattering enhances the detection of vertical
upcoming events as discussed in the next section.

\section{Detection}
\label{sec:det} 

Next generation neutrino telescopes, like IceCube \cite{IceCube} and km3net 
\cite{km3net} will be able to detect these pairs of upward-going charged 
tracks.
The NLSPs/NLKPs production was determined always assuming they will be 
produced below the horizon in relation to the detector. This avoids the large 
background of multiple muons 
produced in cosmic-ray air showers. High $p_T$ muons in air showers could 
mimic widely separated pairs. Isolated downgoing muons have been observed 
as far as 400 m from a shower core \cite{highpt}. 

The minimum detectable separation for charged pairs is an experimental question.  However, a reasonable guess is that the minimum separation for nearly vertical pairs is comparable to the string spacing, 125 m in IceCube.  For pairs that are more horizontal, the smaller separations may be visible, since, in the vertical direction, optical modules are spaced every 17 m, providing finer pixellation.  Here, we will assume that the average minimum separation distance is 100 m.  In km3net, the spacing is unknown, but this is still a reasonable figure. 

A detectable separation of 100~m would eliminate almost completely the
di-muon background. This can be seen in Figure~7 of Ref.~\cite{abcd}. Multiple
scattering will however enhance NLSP/NLKP detection since it enhances the
separation of events near the 100~m threshold for detecting two tracks.

As described in the last section, with the 100 meter minimum distance, multiple 
scattering increases the number of visible 300~GeV (600 and 900~GeV) NLKPs by 46\%
(30 and 21\%). NLSPs will have their number increased by 3\%. Another
feature to be accounted for is the detection dependence on the zenith angle.

The pair separation also depends on the charged particle arrival direction.
Vertical upcoming particles can be produced farther
from the detector. Therefore they have larger separation distances in
the detector and also the multiple scattering effect is larger when
compared to horizontal events. 

Multiple scattering increases the number
of near vertical upcoming NLSPs pairs separated by more than 100~meters by 
24\%. Here, near vertical(horizontal) is defined as pairs within a 5 degrees of
vertical (horizontal) direction. 
For the near horizontal events multiple scattering has an insignificant effect
the increase in the number of NLSPs separated by more then 100~meters is
only 1.8\%.
Although vertical upgoing NLSPs are a tiny fraction of the event rate,
we show the enhancement effect to highlight the fact that
multiple scattering effect increases with increasing zenith angle.
In this way, the far from horizontal events will be more affected by
multiple scattering.

For NLKP pairs multiple scattering is also less effective for near 
vertical events when compared to near horizontal events. As an example,
300~GeV near vertical pairs
will have an 67\% increase in the number of events with more than 100~m
separation when multiple scattering is included while near horizontal
events will have a correspondent 16\% increase.

\section{Conclusions}

We have determined the multiple scattering effect on NLKP and NLSP
particles transversing the Earth. We have shown that multiple 
scattering will enhance the separation between NLKP pairs and between
far from horizontal NLSPs. 

The pair separation in the detector is a clear signature both for NLKPs
as for NLSPs. As discussed in section~\ref{sec:det}, 100~meters separation
is an average lower limit on the experimental separation detectability.
Although this separation by itself gets rid of most of the dimuon
background, it will still enhance the detectability of the next to
lightest particles. 

We find a 46\% increase on the number of 300~GeV NLKPs separated by
more than 100~meters and 30 and 21\% for 600 and 900~GeV NLKPs when
multiple scattering is taken into account.

The effect is less
significant for NLSPs. The number of pairs separated by
more than 100~meters has a 3\% increase when multiple scattering is taken into
account.  Scattering can, however, enhance the
number of NLSPs that are produced far from the horizontal direction. 

In conclusion, NLKP multiple scattering 
through the Earth is an important effect to be considered. This importance of scattering increases for
pairs that are closer to vertical.

This work was partially funded by the U.S. National Science Foundation 
under grant number 0653266, the U.S.
Department of Energy under contract numbers DE-AC-76SF00098 and
DE-AC02-07CH11359 and the Brazilian National Counsel for Scientific
Research (CNPq).


\begin{thebibliography}{99}

\bibitem{abc}
  I.~Albuquerque, G.~Burdman and Z.~Chacko,
  Phys.\ Rev.\ Lett.\  {\bf 92}, 221802 (2004).

\bibitem{abcd}
  I.~Albuquerque, G.~Burdman and Z.~Chacko,
  Phys.\ Rev.\ D\  {\bf 75}, 035006 (2007).

\bibitem{abkn}
I.~Albuquerque, G.~Burdman, C.~Krenke, and B.~Nosratpour,
  Phys.\ Rev.\  D {\bf 78}, 015010 (2008).

\bibitem{madrid} B.~Canadas, D.~G.~Cerdeno, C.~Munoz and S.~Panda,
  arXiv:0812.1067 [hep-ph].

\bibitem{ina}
  M.~H.~Reno, I.~Sarcevic and S.~Su,
  Astropart.\ Phys.\  {\bf 24}, 107 (2005) and
 Y.~Huang, M.~H.~Reno, I.~Sarcevic and J.~Uscinski,
  Phys.\ Rev.\  D {\bf 74}, 115009 (2006).

\bibitem{pdg}C. Amsler {\it et al.}, Phys. Lett. {\bf B667}, 1 (2008).

\bibitem{tsai} Y.~S.~Tsai, Rev. Mod. Phys. {\bf 46}, 815 (1974).

\bibitem{quigg}  R.~Gandhi, C.~Quigg, M.~H.~Reno and I.~Sarcevic,
  Astropart.\ Phys.\  {\bf 5}, 81 (1996)

\bibitem{crust}http://hyperphysics.phy-astr.gsu.edu/Hbase/tables/elabund.html
\bibitem{mantle}http://www.mpch-mainz.mpg.de/~jesnow/Ozeanboden/2001/Lecture4/Composition.html
\bibitem{core}http://en.wikipedia.org/wiki/Earth\#Chemical\_composition

\bibitem{bfield}N. Olsen {\it et al.}, Geophys. J. Int. {\bf 166}, 67 (2006).

\bibitem{IceCube}S. Klein, preprint arXiv:0807.0034.

\bibitem{km3net}C. Markou, Nucl. Instrum. \& Meth. {\bf A595}, 54 (2008). 

\bibitem{highpt}S. Klein and D. Chirkin, in preprint arXiv:0711.0353.
                                             
\end{thebibliography}
\end{document}